\documentstyle[12pt,epsfig]{article}

\begin{document}

\hfill{UB/FI 97-1}\break
\smallskip
\hfill{June 1997}\break

\renewcommand{\thefootnote}{\fnsymbol{footnote}}
\vspace{2 cm}
\centerline{\bf {\LARGE Is the solar neutrino deficit energy-dependent? 
\footnote{Contributed paper to the XVIII International Symposium on 
Lepton-Photon Interactions, Hamburg, Germany, 28 July - 1 
August 1997}}}
\vspace{15 mm}

\centerline{G. Conforto${}^{a,b}$, A. Marchionni${}^{b}$, F. Martelli${}^{a,b}$ 
and F. Vetrano${}^{a,b}$} 

\begin{center}
${}^{a}$ Universit\`a degli Studi, I-61029 Urbino, Italy\\
${}^{b}$ Istituto Nazionale di Fisica Nucleare, I-50125 Firenze, 
Italy
\end{center}
\vspace{2cm}

\begin{abstract}
All existing measurements of the solar 
neutrino flux are compared with the predictions of the most 
recent solar model by Bahcall and Pinsonneault, modified by 
introducing the hypothesis of neutrino oscillations with 
mass differences large enough to render energy-independent 
any quantity observable on earth. It is concluded that the 
data are consistent with this hypothesis and that, at least 
for the time being, any energy-dependence of the solar 
neutrino deficit must be regarded as just an attractive 
theoretical possibility, but not as a compelling reality.
\end{abstract}
\vfill
\eject

\section{Introduction \label{intro}}

	It is by now well established that the measurements of 
the solar neutrino flux on earth fall short of the 
expectations based on solar models.

	The question of whether this deficit can be proven to 
be energy-dependent is of primary importance in neutrino 
physics. In fact, if this were the case and ascribing the 
effect to oscillations, this would imply the existence of a 
$\delta m^{2}$ smaller than $\approx 10^{-5}$ eV${}^{2}$.

	A careful statistical analysis of all the available 
evidence was carried out in ref.~\cite{conf95} with the conclusion 
that the data are consistent with an energy-independent 
depletion of the solar neutrino flux due to vacuum 
oscillations. Other authors \cite{harr95,acke96} have also taken this 
viewpoint.

	However, different results have been obtained in ref.~\cite{krast96} 
 and more recently in ref.~\cite{krast97}. In particular, in this 
last paper, it is claimed that ``an energy independent 
suppression of the solar neutrino spectrum as a result of 
neutrino oscillation/transitions into active neutrinos or 
antineutrinos is ruled out by the data from the four 
operating experiments at 99.96 \% C.L.''

	Two opposite conclusions cannot be both right. To 
investigate the origin of this disagreement, we have 
retraced the steps of the analysis of ref.~\cite{krast97}, starting 
from the same input data detailed in sect.~\ref{Indata}. 

	We can reproduce all results up to the final value of 
the minimum $\chi^2$, but we find that the confidence level for an 
energy-independent depletion of the solar neutrino flux is 
still acceptable, in agreement with our previous results 
\cite{conf95}. A possible explanation for this discrepancy is 
discussed in sect.~\ref{oversight}. We present our final conclusions
in sect.~\ref{conc}.

\section{Input data \label{Indata}}

	In order to make our comparison as tight as possible, 
we have used exactly the same input data as those of ref.~\cite{krast97}. 
The experimental results originate from 
ref~\cite{land96,kirst96,gavr96,fuku96}, the 
predictions of the ``reference'' Standard Solar Model from 
ref.~\cite{bahc95,bahc96}. They are reported in table 1. An update of 
the experimental information is presented in sect.~\ref{conc}.

	The uncertainties on the theoretical predictions are 
obviously correlated. However, at least for the model of 
ref.~\cite{bahc95}, the information available allows \cite{fogl95} to 
calculate the relevant correlation matrix $\rho$ reported in 
table 2.

	Within the approach of ref.~\cite{krast97}, the $\chi^{2}$ for the 
hypothesis of an  oscillation-induced energy-independent 
reduction of all predictions is then simply written as
\[
\chi^{2}(F) = \sum_{i=1}^{4}\sum_{j=1}^{4}\;(e_{i}- t_{i})\; 
(e_{j}-t_{j})\;(S^{-1})_{ij}
\]
where:

\begin{itemize}
\item the indices $i$ and $j$ run over the ``four operating 
experiments'';
\item $e_{i}$ are the experimental results;
\item $t_{i}$ are the theoretical expectations.  For the Gallium 
(Ga), Chlorine (Cl) and Kamiokande (Ka) experiments they 
are obtained from the results $T_{i}$ of the model of 
ref.~\cite{bahc95,bahc96}
through the relations
\[
\begin{array}{rcl}
t_{Ga,Cl} &=& F T_{Ga,Cl} \\
t_{Ka} &=& F (1 - f) T_{Ka} + f T_{Ka}
\end{array}
\] 
where $F$ is the energy-independent depletion factor and $f$ = 
0.155 is the fraction of the Kamiokande detection 
efficiency due to flavour-blind Neutral Currents;

\item $S$ is the covariance matrix. With exactly the same 
procedure as above, the errors on the theoretical 
expectations $\delta_{i}$ are obtained from the estimates of the solar 
model uncertainties $\Delta_{i}$ through the relations
\[
\begin{array}{rcl}
\delta_{Ga,Cl} &=& F \Delta_{Ga,Cl} \\
\delta_{Ka} &=& F (1 - f) \Delta_{Ka} + f \Delta_{Ka}
\end{array}
\] 

They give rise to the solar-model-related part of the 
covariance matrix through the relations $S_{ij} = \rho_{ij} 
\delta_{i} \delta_{j}$ in 
which $\rho_{ij}$ are the elements of the correlation matrix 
reported in table 2. The complete $S$ is then finally 
obtained by adding in quadrature the experimental errors $\sigma_{i}$ 
to the diagonal elements: $S_{ii}^{2} = \delta_{i}^{2} + \sigma_{i}^{2}$. 
The $\sigma_{i}$ are in turn 
obtained by adding in quadrature statistical and 
instrumental errors. When asymmetric errors are given, the 
average values have been used.
\end{itemize}

\section{A trivial oversight ? \label{oversight}}

	The minimum value of the function $\chi^{2}= \chi^{2}(F)$ is 
$\chi^{2}_{\min}$ = 
13.5 corresponding to the result $F = 0.43 \pm 0.06$, in 
excellent agreement with the value $\chi^{2}_{\min}$ = 13.0 quoted in 
ref.~\cite{krast97}.

	The corresponding Confidence Level (CL) for $n$ = 3 (4 
data points minus 1 unknown parameter) degrees of freedom 
is 0.36 \%. Fig. 1 shows the plot of CL = CL($\chi^{2}$) for 
different values of $n$.

	The value $\chi^{2}_{\min}$ = 13.0  found in 
ref.~\cite{krast97} corresponds to 
a CL = 0.46 \%. We find no justification for the low value 
(CL = $4 \times 10^{-4}$) invoked to reject the energy-independent-depletion 
hypothesis.

	A possible explanation that we venture to put forward 
lies in the fact that in Fig. 1 a vertical line at $\chi^{2}$ = 13.0 
intercepts the $n=1$ curve in the range $3 \times 10^{-4} <$ CL $< 4 \times
10^{-4}$. So it is perhaps plausible that the confidence level 
of ref.~\cite{krast97} was calculated for the wrong number of degrees 
of freedom ($n=1$ instead of $n=3$).

\begin{figure}
\begin{center}\mbox{\epsfig{file=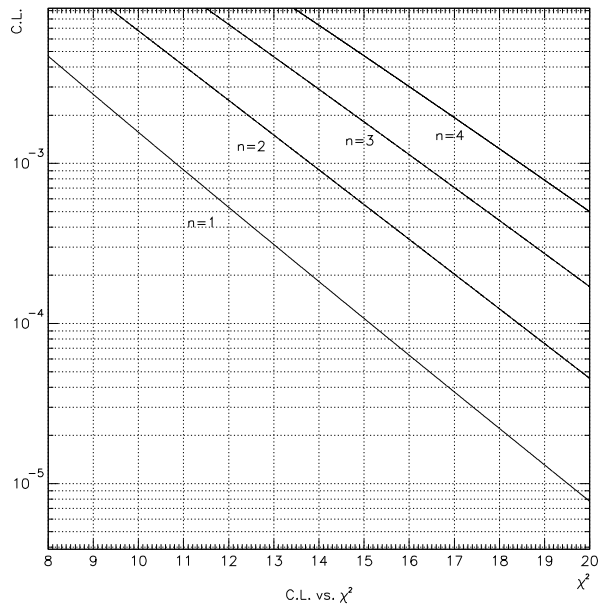,height=9cm,width=9cm}}\end{center}
\caption{Confidence Levels (CL) vs. $\chi^{2}$ for various degrees of 
freedom ($n$).}
\end{figure}

\section{Conclusions \label{conc}}

	The validity of a $\chi^{2}$ analysis relies on the two basic 
hypotheses that errors must be correctly calculated and 
Gaussian-distributed. Neither is really justified in the 
case of solar neutrinos.

	On the experimental side, results are often affected 
by large systematics and asymmetric errors. Occasionally, 
doubts have been raised about the correctness of their 
evaluations.

	On the theoretical side, different models of the same 
sun giving rise to different predictions cannot be all 
simultaneously right, implying that the quoted errors are 
probably underestimated. In some cases the assumption of a 
Gaussian probability density function is highly 
questionable and this in turn implies the likelihood of an 
underestimated variance.

	For all these reasons, the $\chi^{2}$ approach cannot be taken 
too seriously far away from the minimum where small deviations from
a Gaussian behaviour in the tails of the probability density functions
give rise to huge variations of the Confidence Level.

However, even discarding all these caveats, no serious 
disagreement with the energy-independent-depletion hypothesis can
be claimed to exist at present. Making use of the recent and
more precise measurement of the solar neutrino flux by the
SuperKamiokande experiment, $(2.65 \pm 0.15) \times 10^{6}$
cm${}^{-2}$ s${}^{-1}$ \cite{kear97}, yields $\chi^{2}_{\min}$ = 14.2,
corresponding to $F = 0.41 \pm 0.05$ and CL = 0.26 \%. 
The confidence levels are small, but not unacceptably 
small.

	To gauge the acceptability of the CL obtained, it is 
perhaps worth remembering that ``If this probability is 
larger than an agreed-upon  value (0.001, 0.01, or 0.05 are 
common choices), the data are \textit{consistent}  with the 
assumptions; otherwise we may want to find improved 
assumptions. As for the converse, most people do not regard 
a model as truly \textit{inconsistent} unless the probability is as 
low as that corresponding to four or five standard 
deviations for a Gaussian $(6.3\times 10^{-5}$ or $5.7\times 10^{-7})$'' 
\cite{PDG96}.

	The onus of the proof is on the energy-dependence 
camp. As no definitive proof of the correctness of this 
hypothesis has been produced so far, the energy-dependence 
of the solar neutrino deficit must be regarded, at least 
for the time being, as just an attractive theoretical 
possibility, but not as a compelling reality.

\vfill
\eject

\begin{table}
\caption{Solar neutrino experimental results [6-9] and ``reference'' 
Standard Solar Model [10,11] theoretical predictions.}
\vspace{0.5cm}
\begin{center}
\begin{tabular}
              {lcc}
Experiment (Units)	&	Result	&	Prediction	\\
\hline
 & & \\
Gallex (SNU)    &  $69.7\pm 6.7^{+3.9}_{-4.5}$  &   \\
                &                               & $136.8^{+8}_{-7}$	\\
Sage (SNU)      &  $72^{+12}_{-10}{}^{+5}_{-7}$ & 	\\
 & & \\
Chlorine (SNU)  &  $2.56\pm 0.16\pm 0.14$       & $9.5^{+1.2}_{-1.4}$	\\
 & & \\
Kamiokande ($10^{6}$cm${}^{-2}$s${}^{-1}$)
			&	$2.80\pm 0.19\pm 0.33$
			&	$6.62^{+0.93}_{-1.12}$ \\

\end{tabular}
\end{center}
\end{table}

\begin{table}
\caption{Error correlation matrix of the theoretical predictions for 
the Gallium (Ga), Chlorine (Cl) and Kamiokande (Ka) 
experiments.
The convention for the indices of the matrix elements $\rho_{ij}$ 
is 
1=Ga, 2=Cl, 3=Ka.}
\vspace{0.5cm}
\begin{center}
\begin{tabular}{ccccc}
	&	&	1	& 0.656 & 0.646 \\
$\rho$	&   =	&      0.656	& 1	& 0.976 \\
	&	&      0.646	& 0.976 & 1
\end{tabular}
\end{center}
\end{table}


\begin{thebibliography}{99}


\bibitem{conf95} G. Conforto, Nucl. Phys. B (Proc. Suppl.) {\bf 38}, 308 
(1995). Also, G. Conforto {\it et al.}, Astropart. Phys. {\bf 5}, 147 (1996), 
hep-ph/9606226.

\bibitem{harr95} P.F. Harrison, D.H. Perkins, W.G. Scott, Phys. Lett. 
{\bf B349}, 137 (1995); Phys. Lett., {\bf B396}, 186 (1997), hep-ph/9702243.

\bibitem{acke96} A. Acker, S. Pakvasa, Phys. Lett. {\bf B397}, 209 (1997), 
hep-ph/9611423. 

\bibitem{krast96} P.I. Krastev and S.T. Petcov, Phys. Rev. {\bf D53}, 1665 
(1996), hep-ph/9510367.

\bibitem{krast97} P.I. Krastev and S.T. Petcov, Phys. Lett. {\bf B395}, 69 
(1997), hep-ph/9612243.

\bibitem{land96} K.Lande, to appear in Proceedings of Neutrino 
'96 Int. Conference, June 13-19 1996, Helsinki.

\bibitem{kirst96} T. Kirsten, to appear in Proceedings of  
Neutrino '96 Int. Conference, June 13-19 1996, Helsinki.

\bibitem{gavr96} V. Gavrin, to appear in Proceedings of  
Neutrino  '96 Int. Conference, June 13-19 1996, Helsinki. 

\bibitem{fuku96} Y. Fukuda {\it et al.}, Phys. Rev. Lett. {\bf 77}, 1683 (1996).

\bibitem{bahc95} J. Bahcall and M.H. Pinsonneault, Rev. Mod. Phys. {\bf 67}, 
781 (1995).

\bibitem{bahc96} J. Bahcall {\it et al.}, Phys. Rev. {\bf C54}, 411 (1996), 
nucl-th/9601044.

\bibitem{fogl95} G.L. Fogli and E. Lisi, Astropart. Phys. {\bf 3}, 185 (1995).

\bibitem{kear97} E. Kearns, talk given at the ``Workshop on fixed
target physics at the Main Injector'', May 1-4 1997, FNAL, Batavia,
Illinois.

\bibitem{PDG96} Particle Data Group, R. M. Barnett {\it et al.}, Phys. Rev. 
{\bf D54}, 1 (1996).

\end{thebibliography}
\end{document}